\documentstyle[12pt,epsf,rotating]{article}

\catcode`@=11
\def\citer{\@ifnextchar [{\@tempswatrue\@citexr}{\@tempswafalse\@citexr[]}}
 
\def\@citexr[#1]#2{\if@filesw\immediate\write\@auxout{\string\citation{#2}}\fi
  \def\@citea{}\@cite{\@for\@citeb:=#2\do
    {\@citea\def\@citea{--\penalty\@m}\@ifundefined
       {b@\@citeb}{{\bf ?}\@warning
       {Citation `\@citeb' on page \thepage \space undefined}}%
\hbox{\csname b@\@citeb\endcsname}}}{#1}}
\catcode`@=12
 
\newcommand{\lsim}{\raisebox{-0.13cm}{~\shortstack{$<$ \\[-0.07cm] $\sim$}}~}
\newcommand{\gsim}{\raisebox{-0.13cm}{~\shortstack{$>$ \\[-0.07cm] $\sim$}}~}

\newcommand{\beq}{\begin{equation}}
\newcommand{\eeq}{\end{equation}}
\newcommand{\bea}{\begin{eqnarray}}
\newcommand{\eea}{\end{eqnarray}}
\newcommand{\CP}{\mbox{${\cal CP}$}}

\newcommand{\tgb}{\mbox{tg$\beta$}}
\newcommand{\tb}{\mbox{tg$\beta$}}

\newcommand{\ctowidth}[2]{ \setbox\mycount=\hbox{$#2$}
                          \hbox to \wd\mycount{$ \hss #1 \hss $} }
\newcommand{\ltowidth}[2]{ \setbox\mycount=\hbox{$#2$}
                          \hbox to \wd\mycount{$\hskip0pt plus0pt minus1fil
                           #1 \hfill $} }
\newcommand{\rtowidth}[2]{ \setbox\mycount=\hbox{$#2$}
                          \hbox to \wd\mycount{$\hfill #1
                          \hskip0pt plus0pt minus1fil$} }

\begin{document}

\renewcommand{\thefootnote}{\fnsymbol{footnote}}
\setcounter{page}{0}

\begin{titlepage}

\begin{flushright}
DESY 98-159 \\
hep-ph/9810289 \\
October 1998
\end{flushright}

\vspace*{1cm}

\begin{center}
{\large \sc Higgs Boson Production and Decay at the
Tevatron\footnote{Contribution to the Workshop {\it Physics at Run II --
Supersymmetry/Higgs}, 1998, Fermilab, USA}
}

\vspace*{1cm}

{\sc Michael Spira}

\vspace*{1cm}

{\it II.\ Institut f\"ur Theoretische Physik\footnote{Supported by
Bundesministerium f\"ur Bildung und Forschung (BMBF), Bonn, Germany,
under Contract 05~7~HH~92P~(5), and by EU Program {\it Human Capital and
Mobility} through Network {\it Physics at High Energy Colliders} under Contract
CHRX--CT93--0357 (DG12 COMA).}, Universit\"at Hamburg, Luruper Chaussee
149, D--22761 Hamburg, Germany}

\end{center}

\vspace*{2cm}

\begin{abstract}
\normalsize
\noindent
The theoretical status of Higgs boson production and decay at the Tevatron
within the Standard Model and its minimal supersymmetric extension is
reviewed. The focus will be on the evaluation of higher-order corrections
to the production cross sections and their phenomenological implications.
\end{abstract}

\end{titlepage}

\renewcommand{\thefootnote}{\arabic{footnote}}

\setcounter{footnote}{0}
\setcounter{page}{2}

\section{Introduction}
The search for Higgs bosons is of primary interest for present and
future experiments. In the Standard Model [SM] the existence of a single
Higgs particle
is predicted as a consequence of electroweak symmetry breaking by means
of the Higgs mechanism \cite{1}. The direct search at the LEP experiments leads
to a lower bound of $\sim 90$ GeV \cite{2} on the value of the Higgs mass, while
triviality and unitarity constraints require the Higgs mass to be
smaller than ${\cal O}$(1~TeV). At the upgraded Tevatron experiment, a
$p\bar p$ collider with a c.m.\ energy $\sqrt{s}=2$ TeV, SM Higgs bosons
will mainly be produced via gluon fusion $gg\to H$ and the
Drell--Yan like production $q\bar q \to W^* \to WH$. Since an
intermediate mass Higgs boson will dominantly decay into $b\bar b$
pairs, the QCD background of $b\bar b$ production will be too large to
allow for a detection of the Higgs boson produced in the gluon fusion
process.
Recently it has been shown that a detection of the Higgs boson from $W$
fusion $WW\to H$ seems to be impossible due to the overwhelming QCD
background \cite{31}.
Thus, the primary possibility to find the SM Higgs boson at the
Tevatron will be via the Drell--Yan like process. Careful studies have
shown that a discovery of the SM Higgs boson at the upgraded Tevatron
might be possible for Higgs masses up to about 120 GeV \cite{3}. Apart from the
dominant $b\bar b$ decay \cite{3} it turned out that a discovery may also be
feasible via the $H\to \tau^+\tau^-$ decay \cite{4} in $H+W/Z$ production, while
the gold-plated mode of the LHC, $gg\to H\to ZZ^*\to 4\ell^\pm$, is
hopeless at the Tevatron \cite{3}. Recently it has been found that there
is also the possibility to detect the processes $gg\to H\to W^*W^* \to
\ell \nu jj, \ell^+\ell^- \nu \bar\nu$ by using the strong angular
correlations among the final state leptons \cite{5}.

The SM Higgs mass is unstable against quantum fluctuations in the
context of grand unified theories, which force the Higgs mass to be of
the order of the GUT scale $M_{GUT} \sim 10^{16}$ GeV. The most
attractive solution to this hierarchy problem is provided by the
introduction of supersymmetry. The minimal supersymmetric extension of
the SM [MSSM] requires two Higgs doublets, leading
to the existence of 5 elementary Higgs particle, two neutral \CP-even
($h,H$), one neutral \CP-odd ($A$) and two charged ones ($H^\pm$). At
tree-level the mass of the light scalar Higgs boson $h$ is restricted to be
smaller than the $Z$ mass $M_Z$. However, radiative corrections to
the MSSM Higgs sector are large, since they increase with the fourth
power of the large top quark mass $m_t$. They enhance the upper bound on
the light scalar Higgs mass to $M_h\lsim 135$ GeV \cite{7}.

For the discovery at the Tevatron the light scalar Higgs boson $h$ will
mainly be produced via $q\bar q\to
W/Z + h$ analogously to the SM case. However, for large $\tgb$ the
associated production meachanisms $q\bar q, gg\to b\bar b + h/A$ become
competitive due to the enhanced Yukawa couplings to $b$
quarks. Finally, similar to the LHC the light scalar Higgs may be
detectable in SUSY particle production process via the decay
$\tilde \chi_2^0\to \tilde \chi_1^0 h$ in the final state cascade
decays \cite{8}.

Charged Higgs bosons can also be searched for at the Tevatron \cite{roy}.
They will be produced in top quark decays $t\to bH^+$,
if their masses are light anough. At the Tevatron the process $p\bar p
\to t\bar t$ with $t\to bH^+$ is sensitive to charged Higgs bosons for
large $\tgb$. The Drell--Yan like charged Higgs pair
production $p\bar p \to H^+ H^-$ and gluon fusion $gg\to H^+H^-$
processes seem to be difficult to detect,
while the analysis of the associated production $p\bar p\to W^\pm H^\mp$
requires a more careful background study in order to investigate its
relevance for charged Higgs searches at the Tevatron. Finally
gluon-gluon fusion $gg\to t\bar b H^-$, gluon-bottom fusion $gb \to t
H^-$ and quark-bottom fusion $qb\to q'bH^\pm$ provide additional
possibilities to search for charged Higgs bosons at the Tevatron.

\section{Higgs Boson Decays}

\begin{figure}[hbtp]
\vspace*{-0.3cm}

\hspace*{2.5cm}
\begin{turn}{-90}%
\epsfxsize=7cm \epsfbox{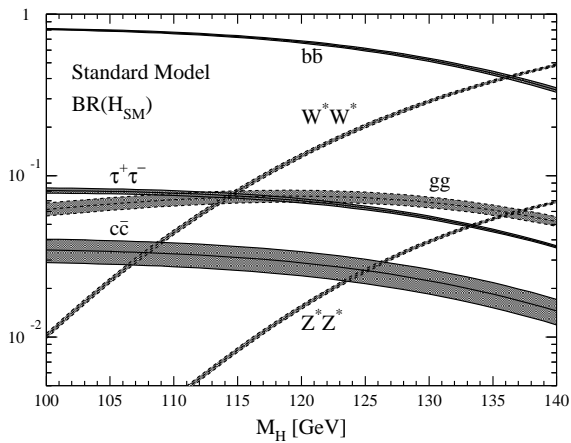}
\end{turn}
\vspace*{-1.0cm}

\caption[ ]{\label{fg:1} \it Branching ratios of the
dominant decay modes of the SM Higgs particle. All relevant higher order
corrections are taken into account. The shaded bands represent the
variations due to the uncertainties in the input parameters:
$\alpha_s(M_Z^2) = 0.120 \pm 0.003$, $\overline{m}_b(M_b) = (4.22 \pm 0.05)$
GeV, $\overline{m}_c(M_c) = (1.22 \pm 0.06)$ GeV, $M_t = (174 \pm 5)$ GeV.}
\end{figure}

\noindent
\underline{\it $\phi\to f\bar f$} \\[0.2cm]
In the intermediate Higgs mass range the SM Higgs decay $H\to b\bar b$
is dominating, while the decay $H\to \tau^+\tau^-$ reaches a branching
ratio of ${\cal O}(10\%)$, see Fig.~\ref{fg:1}.
In the past the QCD corrections have been evaluated
\cite{9}. They turn out
to be moderate for the decay mode $H\to t\bar t$ in the threshold
region, while they are large for the $H\to b\bar b, c\bar c$ decays due
to large logarithmic contributions. These can be absorbed in the running
quark mass by replacing the scale by the Higgs mass. In order to gain a
consistent prediction of the partial decay widths one has to use direct
fits of the $\overline{\rm MS}$ masses $\overline{m}_Q(M_Q)$ to
experimental data. The evolution of $\overline{m}_Q(M_Q)$ to
$\overline{m}_Q(M_H)$ is controlled by the renormalization
group equations for the running $\overline{\rm MS}$ masses. As a result the QCD
corrections reduce the $H\to b\bar b$ decay width by $\sim 50\%$ and the
$H\to c\bar c$ width by $\sim 75\%$ relative to the Born term expressed
in terms of the quark pole masses $M_Q$.

In the MSSM the decay modes $h,H,A\to b\bar b, \tau^+\tau^-$ dominate
the neutral Higgs decay modes for large $\tgb$, while for small $\tgb$
they are important for $M_{h,H,A}\lsim 150$ GeV as can be inferred from
Figs.~\ref{fg:2}a--c.
The dominant decay
modes of the charged Higgs particles are $H^+\to \tau^+\nu_\tau, t\bar
b$ [see Fig.~\ref{fg:2}d]. Analogous to the SM case the QCD corrections
reduce the partial
decay widths into $b,c$ quarks by about 50--75\% as a result of the
running quark masses, while they are moderate for decays into top
quarks.
\begin{figure}[hbt]
\vspace*{-1.2cm}

\hspace*{-2.5cm}
\begin{turn}{-90}%
\epsfxsize=8cm \epsfbox{br.l}
\end{turn}
\vspace*{-1.8cm}

\centerline{\bf Fig.~\ref{fg:2}a \hspace*{6.0cm} Fig.~\ref{fg:2}b}

\vspace*{-7.05cm}

\hspace*{5.5cm}
\begin{turn}{-90}%
\epsfxsize=8cm \epsfbox{br.h}
\end{turn}
\vspace*{-1.8cm}


\vspace*{-1.0cm}

\hspace*{-2.5cm}
\begin{turn}{-90}%
\epsfxsize=8cm \epsfbox{br.a}
\end{turn}
\vspace*{-1.8cm}

\centerline{\bf Fig.~\ref{fg:2}c \hspace*{6.0cm} Fig.~\ref{fg:2}d}

\vspace*{-7.05cm}

\hspace*{5.5cm}
\begin{turn}{-90}%
\epsfxsize=8cm \epsfbox{br.c}
\end{turn}
\vspace*{-1.5cm}


\caption[]{\label{fg:2} \it Branching ratios of the MSSM Higgs bosons $h
(a), H (b), A (c), H^\pm (d)$ for non-SUSY decay modes as a function of their
masses for two values of $\tb=3, 30$ and vanishing mixing. The common squark
mass has been chosen as $M_S=1$ TeV.}
\end{figure}

Below the corresponding thresholds the decays $A\to t^*t$ and $H^+\to
t^*\bar b$ into off-shell top quarks turn out to be important, since
they reach branching ratios of ${\cal O}(1\%)$ already far below the
thresholds for on-shell top quarks \cite{13}. \\

\noindent
\underline{$H\to WW,ZZ$} \\[0.2cm]
In the SM the decays $H\to WW,ZZ$ are dominant for $M_H\gsim 140$ GeV,
since they increase with the third power of the Higgs mass for large
Higgs masses, see Fig.~\ref{fg:1}.
Decays into off-shell $W,Z$ bosons $H\to
W^*W^*,Z^*Z^*$ are sizeable already for Higgs masses $M_H\gsim 100$ GeV,
i.e.\ significantly below the $WW/ZZ$ thresholds \cite{13,16}.

In the MSSM the decays $h,H\to WW,ZZ$ are suppressed by kinematics and,
in general, by SUSY couplings and are thus less important than in the
SM. Their branching ratios turn out to be sizeable only for small $\tgb$
or in the decoupling regime, where the light scalar Higgs particle $h$
reaches the upper bound of its mass.

The gluonic (photonic) decays of the Higgs bosons $h,H,A\to gg (\gamma
\gamma)$ reach branching ratios of $\sim 10\%$ ($\sim 10^{-3}$) in the
SM and MSSM and are unimportant at the Tevatron. \\

\noindent
\underline{$H\to hh, AA$} \\[0.2cm]
The decay mode $H\to hh$ is dominant in the MSSM for small $\tgb$ in a
wide range of heavy scalar Higgs masses $M_H$ below the $t\bar t$
threshold, see Fig.~\ref{fg:2}b. The dominant radiative corrections to
this decay arise from
the corrections to the self-interaction $\lambda_{Hhh}$ in the
MSSM and are large \cite{7}. The decay mode $H\to AA$ is only important at the
lower end of the heavy scalar Higgs mass range. \\

\noindent
\underline{$\phi\to \phi' + W/Z$} \\[0.2cm]
The decay modes $H\to AZ$ and $A\to hZ$ are important for small $\tgb$
below the $t\bar t$ threshold, see Figs.~\ref{fg:2}b,c. Similarly the
decays $H^+\to W^*A,W^{(*)}h$ are sizeable
for small $\tgb$ and $M_{H^+}< m_t+m_b$, see Fig.~\ref{fg:2}d. The dominant
higher-order corrections
can be absorbed into the couplings and masses of the Higgs sector. Below the
corresponding thresholds decays into off-shell Higgs and gauge bosons turn out
to be important especially for small $\tgb$ \cite{13}, see Figs.~\ref{fg:2}b--d.
\\

\begin{figure}[hbt]
\vspace*{0.2cm}

\hspace*{3.0cm}
\begin{turn}{-90}%
\epsfxsize=6cm \epsfbox{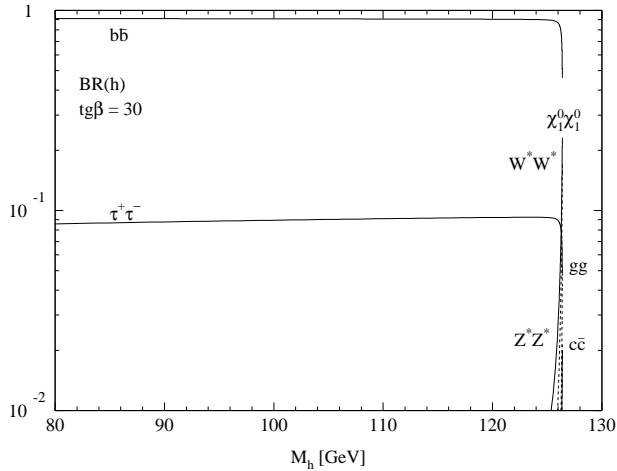}
\end{turn}
\vspace*{-0.0cm}

\caption[]{\label{fg:3} \it Branching ratios of the light scalar MSSM Higgs
boson $h$ decays into charginos/neutralinos as a function of its mass for
$\tb=30$. The mixing parameters have been chosen as $\mu=150$ GeV,
$A_t=2.45$ TeV, $A_b=0$ and the squark masses of the first two generations as
$M_{\widetilde{Q}}=1000$ GeV. The gaugino mass parameter has been set to
$M_2=136$ GeV. The masses of the lightest gauginos are $m_{\tilde
\chi_1^0} = 56.5$ GeV and $m_{\tilde \chi_1^\pm} = 94.1$ GeV.}
\end{figure}

\noindent
\underline{\it Decays into SUSY particles} \\[0.2cm]
Higgs decays into
charginos, neutralinos and third-generation sfermions can become
important, once they are kinematically allowed \cite{13a} as can be
inferred from Fig.~\ref{fg:3}, which shows the branching ratios of the
heavy Higgs bosons into gauginos and squarks as functions of their
masses for a special SUSY scenario. Thus they could
complicate the Higgs search at the Tevatron, since the decay into the LSP
will be invisible.

\section{Higgs Boson Production}

\begin{figure}[hbt]
\vspace*{0.2cm}

\hspace*{2.5cm}
\begin{turn}{-90}%
\epsfxsize=7cm \epsfbox{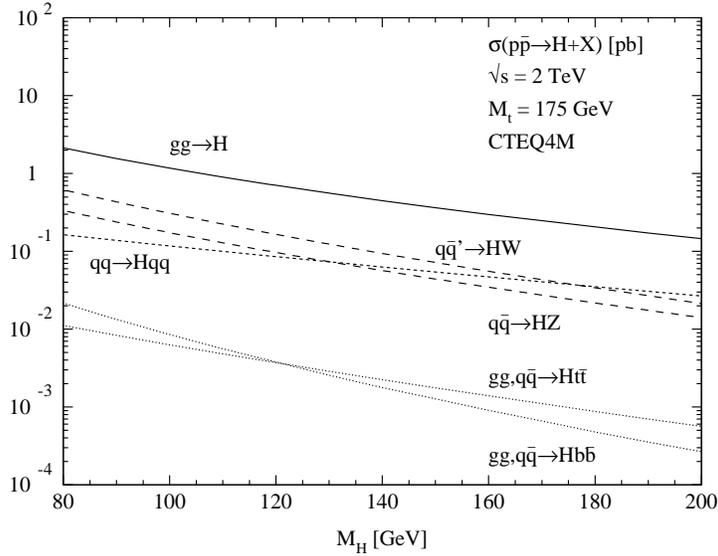}
\end{turn}
\vspace*{-0.2cm}

\caption[]{\label{fg:4} \it Higgs production cross sections at the Tevatron
[$\sqrt{s}=2$ TeV] for the various production mechanisms as a function of the
Higgs mass. The full QCD-corrected results for the gluon fusion $gg
\to H$, vector boson fusion $qq\to VVqq \to Hqq$, Higgs-strahlung
$q\bar q \to V^* \to HV$ and associated production $gg,q\bar q \to Ht\bar t,
Hb\bar b$ are shown.
The QCD corrections to the last process are unknown and thus not included.}
\end{figure}

\noindent
\underline{$q\bar q\to V^*\to VH$ $[V=W,Z]$} \\[0.2cm]
The most relevant SM Higgs production mechanism at the Tevatron is
Higgs-strahlung off $W,Z$ bosons $q\bar q\to W^*/Z^* \to W/Z + H$. The
cross section reaches values of $10^{-1}$--1 $pb$ in the relevant Higgs
mass range $M_H\lsim 120$ GeV, where this production process may be
visible at the Tevatron \cite{3}, see Fig.~\ref{fg:4}.
The QCD corrections coincide with those of the
Drell-Yan process and increase the cross sections by about 30\%
\cite{23,23a}. The
theoretical uncertainty can be estimated as $\sim 15\%$ from the
remaining scale dependence. The dependence on different sets of parton
densities is rather weak and leads to a variation of the
production cross sections by about 15\%.

In the MSSM the Higgs-strahlung processes are in general suppressed by
the SUSY couplings. However, the process $q\bar q\to W^*/Z^*\to W/Z + h$
can be important in the decoupling regime, where the light scalar Higgs
particle $h$ exhibits SM Higgs properties, see Fig.~\ref{fg:5}a,b. \\
\begin{figure}[hbtp]
\vspace*{0.3cm}

\hspace*{2.3cm}
\begin{turn}{-90}%
\epsfxsize=7cm \epsfbox{cxn.h3}
\end{turn}
\vspace*{0.3cm}

\centerline{\bf Fig.~\ref{fg:5}a}

\vspace*{0.2cm}

\hspace*{2.3cm}
\begin{turn}{-90}%
\epsfxsize=7cm \epsfbox{cxn.h30}
\end{turn}
\vspace*{0.3cm}

\centerline{\bf Fig.~\ref{fg:5}b}

\caption[]{\label{fg:5} \it Neutral MSSM Higgs production cross
sections at the Tevatron [$\sqrt{s}=2$ TeV] for gluon fusion $gg\to \Phi$,
vector-boson fusion $qq\to qqVV \to qqh/
qqH$, vector-boson bremsstrahlung $q\bar q\to V^* \to hV/HV$ and the associated
production $gg,q\bar q \to \Phi b\bar b/ \Phi t\bar t$ including all known
QCD corrections. (a) $h,H$ production for $\tb=3$, (b) $h,H$ production for
$\tb=30$, (c) $A$ production for $\tb=3$, (d) $A$ production for $\tb=30$.}
\end{figure}
\addtocounter{figure}{-1}
\begin{figure}[hbtp]
\vspace*{0.5cm}

\hspace*{2.2cm}
\begin{turn}{-90}%
\epsfxsize=7.0cm \epsfbox{cxn.a3}
\end{turn}
\vspace*{0.5cm}

\centerline{\bf Fig.~\ref{fg:5}c}

\vspace*{0.5cm}

\hspace*{2.2cm}
\begin{turn}{-90}%
\epsfxsize=7.0cm \epsfbox{cxn.a30}
\end{turn}
\vspace*{0.5cm}

\centerline{\bf Fig.~\ref{fg:5}d}

\caption[]{\it Continued.}
\end{figure}

\noindent
\underline{$q\bar q, gg\to \phi t\bar t, \phi b\bar b$} \\[0.2cm]
In the SM both processes of Higgs radiation off top and bottom quarks
are unimportant due to the small event rates, see Fig.~\ref{fg:4}. However,
in the MSSM Higgs
radiation off bottom quarks becomes important for large $\tgb$ with cross
sections exceeding 10 $pb$ for the light scalar ($h$) and the
pseudoscalar ($A$) Higgs particles, see Fig.~\ref{fg:5}. Thus, the
theoretical prediction is
crucial for the large $\tgb$ regime in the MSSM.

Until now the full QCD
corrections are unknown so that the cross sections are only known within
about a factor of 2. However, the QCD corrections are known in the limit
of light Higgs particles compared with the heavy quark mass, which is
applicable for $t\bar t +h$ production \cite{24}. In this limit the cross
section
factorizes into the production of a $t\bar t$ pair, which is convoluted
with a splitting function for Higgs radiation $t\to t+h$. This reults in
an increase of the cross section by about 20--60\%. However, since this
equivalent Higgs approxiamtion is only valid within a factor of 2 the
result may not be sufficiently reliable. Moreover, it is not valid for
bottom quarks, which are more relevant for the Tevatron.

In the opposite limit of large Higgs masses $M_H\gg M_Q$ large
logarithms $\log M^2_H/M^2_Q$ arise due to quasi-on-shell $t$-channel quark
exchanges, which can be resummed by absorbing them into the heavy quark
parton densities. After adding the processes $q\bar q,gg\to b\bar b +
\phi$, $gb \to b + \phi$ and $b\bar b\to \phi$, after the logarithms have
been subtracted accordingly, the final result turns out to be about a
factor of 2 larger than the pure $q\bar q,gg \to b\bar b + \phi$
processes \cite{25}. However, there are additional sources of large logarithms,
e.g.\ Sudakov logarithms from soft gluon radiation and large logarithms
related to the Yukawa coupling, which will appear in the NLO cross
section. Thus, a full NLO calculation is needed in order to gain a
satisfactory theoretical prediction of these production processes. In
this analysis the scales of the parton densities have been identified
with the invariant mass of the final state $Q\bar Q \phi$ triplet and
the bottom Yukawa coupling in terms of the bottom pole mass $M_b = 5$
GeV. \\

\noindent
\underline{$gg\to \phi$} \\[0.2cm]
The gluon fusion processes are
mediated by heavy top and bottom quark triangle loops and in the MSSM by
squark loops in addition \citer{29a,21}. Gluon fusion is the dominant neutral
Higgs
production mechanism at the Tevatron, even for large $\tgb$, with cross
sections of $1$--$10^3 pb$, see Figs.~\ref{fg:4}, \ref{fg:5}. However, the
dominant $b\bar b$ final states are
overwhelmed by the huge QCD background of $b\bar b$ production and thus
hopeless for a detection of the Higgs bosons via gluon fusion. Only the
$\tau^+\tau^-$ decay modes may be promising for large $\tgb$, especially
if the Higgs bosons are produced in association with a jet \cite{26}. Moreover,
similar to the LHC it may be possible to detect the $H\to W^*W^*$ decay
mode in a significant Higgs mass range due to the strong
angular correlations of the final state leptons \cite{5}.

The two-loop QCD corrections enhance the gluon fusion cross sections by
about 60--100\% and are thus large \cite{19}. They are dominated by
soft and collinear gluon radiation in the SM and for small $\tgb$ in the MSSM
\cite{29}.
The remaining scale dependence yields an estimate of $\sim 15\%$ for the
theoretical uncertainty. The dependence of the gluon fusion cross
section on different parton densities amounts to about 15\%. The $K$ factor
remains nearly the same after including squark loops, since the dominant
soft and collinear gluon effects are universal, thus suppressing the (s)quark
mass dependence \cite{21}.

Recently the analytical QCD corrections to Higgs + jet production have
been evaluated in the limit of heavy top quarks, but there is no
numerical analysis so far \cite{30}. \\

\newpage
\noindent
\underline{$WW/ZZ\to H$} \\[0.2cm]
Vector boson fusion $WW/ZZ\to H$ in the SM leads to a sizeable cross
section at the Tevatron, see Fig.~\ref{fg:4}. Since there are two forward
jets in the full
process $qq \to WW/ZZ + qq \to H + qq$, one may hope to be able to suppress
the QCD backgrounds by appropriate cuts. Unfortunatly it turned out that
this is not possible \cite{31}.

The QCD corrections can easily be obtained
in the structure function approach from deep inelastic scattering. They
are small enhancing the cross
section by about 10\% \cite{23a,32}. In the MSSM vector boson fusion is
additionally suppressed by SUSY couplings and thus unimportant at the
Tevatron. \\

\begin{figure}[hbt]
\vspace*{-0.0cm}

\hspace*{2.5cm}
\begin{turn}{-90}%
\epsfxsize=7cm \epsfbox{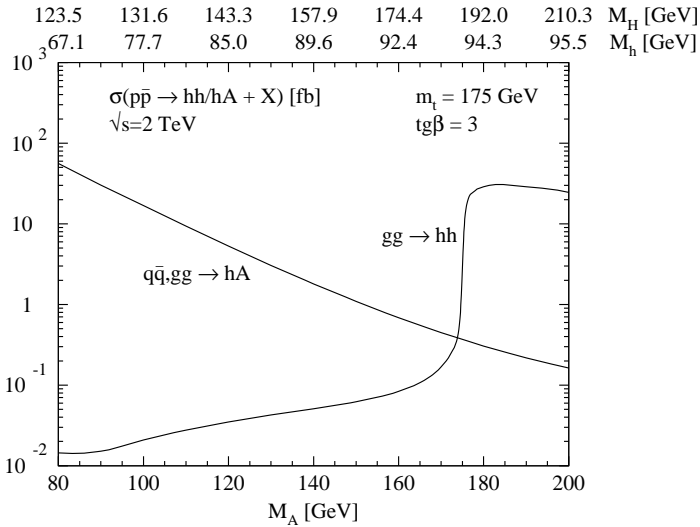}
\end{turn}
\vspace*{0.0cm}

\caption[]{\label{fg:hpair} \it QCD corrected production cross sections of
$hh, hA$ pairs at the Tevatron [$\sqrt{s}=2$ TeV] as a function of the
pseudoscalar Higgs mass for $\tgb = 3$. The secondary axis exhibits the
corresponding values of the light and heavy scalar Higgs masses $M_h,
M_H$. The bump in the $gg\to hh$ cross section originates from resonance
$gg\to H\to hh$ production.}
\end{figure}

\noindent
\underline{\it Higgs pairs} \\[0.2cm]
Light scalar Higgs pair production $gg\to hh$ yields a sizeable cross
section at the Tevatron with $\sigma \gsim 10 fb$ \cite{33,34}.
The cross section for $q\bar q, gg\to hA$ is of similar size in some
regions of the MSSM paramater space [see Fig.~\ref{fg:hpair}]. Since
the process $gg\to H\to hh$ is sensitive to the trilinear
coupling $\lambda_{Hhh}$ it is important for a partial reconstruction of
the Higgs potential. One may hope that the dominant $b\bar bb\bar b,
b\bar b\tau^+\tau^-$ final states can be extracted from the QCD
backgrounds due to the different event topologies. The two-loop QCD
corrections have recently been calculated [for $gg$ initial states in
the limit of heavy top quarks, thus leading to a reliable result for
small $\tgb$]. They enhance the $gg\to hh, hA$ production cross
sections by about 70--90\% and the Drell--Yan-like $q\bar q \to
hA$ cross section by about 30\% \cite{34}.

\section{Conclusions}
All relevant NLO QCD corrections to Higgs boson decays within the SM and
MSSM are known so that the theoretical uncertainties are sufficiently
small. At the Tevatron the decay modes $\phi^0\to b\bar b,
\tau^+\tau^-,W^*W^*$ are relevant for the Higgs boson search in Run II.
All corrections beyond LO are contained in the program HDECAY\footnote{The
computer program is available from http://wwwcn.cern.ch/$\sim$mspira/.}
\cite{23a,35}, which calculates the branching ratios of SM and MSSM Higgs
bosons.

For neutral Higgs boson production most QCD corrections are known
leading to a nearly complete theoretical status. The only processes,
which are known at LO, are Higgs radiation off top and bottom quarks, the
latter being important for large $\tgb$ in the MSSM. The known corrections to
the important processes are moderate and thus well under control. The
remaining theoretical uncertainties are less than $\sim 15\%$\footnote{The
complete computer program library for all production processes of SM and
neutral MSSM Higgs bosons for public use is available from
http://wwwcn.cern.ch/$\sim$mspira/.}. \\

\noindent
{\bf Acknowledgements.} \\
I would like to thank S.\ Dawson, S.\ Dittmaier, A.\ Djouadi, J.\ Kalinowski,
T.\ Plehn and P.M.\ Zerwas for a pleasant collaboration in the topics
presented in this paper and many useful discussions.

\end{document}